\documentclass[12pt]{article}
\usepackage{amsfonts,amsmath,amssymb}
\usepackage{amsthm,amscd,ulem}
\makeatletter

\@addtoreset{equation}{section}
\makeatother
\everymath{\displaystyle}
\theoremstyle{definition}

\def\b1{{ {\bf 1}}}

\begin{document}
\title{
{\bf Non-associative Gauge Theory}}
\author{Takayoshi Ootsuka~$^{\rm a}$\thanks{
e-mail: ootsuka@cosmos.phys.ocha.ac.jp},~
Erico Tanaka~$^{\rm a}$\thanks{
e-mail: erico@cosmos.phys.ocha.ac.jp}~
and Eugene Loginov~$^{\rm b}$\thanks{
e-mail: loginov@ivanovo.ac.ru}
\\
\quad \\ 
{\small $^{\rm a}$Physics Department, Ochanomizu University, 
2-1-1 Bunkyo Tokyo, Japan} \\
{\small $^{\rm b}$Department of Physics, Ivanovo State University, 
Ermaka St. 39, Ivanovo, 153025, Russia}}
\date{}
\maketitle
\begin{abstract}
We present a construction of gauge theory which its structure group is 
not a Lie group, but a Moufang loop which is essentially non-associative.
As an example of non-associative algebra, 
we take octonions with norm one as a Moufang loop, 
with which we can produce an octonionic gauge theory.
Our octonionic gauge theory is a natural generalization
of Maxwell $U(1)\simeq S^1$ gauge theory and Yang-Mills $SU(2)\simeq S^3$
gauge theory.
We also give the BPST like instanton solution 
of our octonionic gauge theory in 8 dimension.
\end{abstract}

\section{Introduction}

The application of non-associative algebra to physics were extremely
rare but not none,
and among them octonions were comparatively well studied.
Octonion is a last member of the normed division algebra, 
which its sequence consists of 
${\Bbb R} \to {\Bbb C} \to {\Bbb H} \to {\Bbb O}$.  
It is a non-commutative and also non-associative algebra~\cite{Baez}.

For the example of applications, octonions were used in the research of 
quark's color symmetry and strong interaction~\cite{G-T}.
They were also used for the construction of higher dimensional
instantons, 
such as 7 and 8 dimensional Yang-Mills instantons~\cite{CDFN}
and gravitational instantons~\cite{B-S, G-P-P, A-O, BFK}.
The 7-dimensional gravitational instanton, a $G_2$ manifold, 
is considered within the compactification program 
of 11-dimensional M-theory~\cite{A-W, A-G}.
In mathematics, octonions were used for constructing exceptional groups 
which are considered frequently in the field of particle
physics~\cite{G-T, Ramond}, 
because of its relation with GUT, supersymmetry and superstring theory.

However, most of these works only use octonions as an auxiliary tool.
There are some works which regard the properties of octonions 
and attempts to incorporate its peculiar feature 
in the construction of the theory, 
for example, such as ``octonionic gauge theory''~\cite{Loginov, OTU}. 
Though these works might have their own importance, 
we think other constitution is needed, 
since they do not have octonionic gauge symmetry as a internal symmetry, 
that is to say,
they are not the natural extension of Maxwell and Yang-Mills gauge theory.
So here we might present our naive motivation.
We have already met a gauge theory, 
a Maxwell gauge theory and Yang-Mills gauge theory.
Maxwell gauge theory is abelian, and has $U(1)$ gauge symmetry.
Since $U(1)$ is a set of complex numbers of norm one, 
it could be thought as $S^1$.
So we may call the Maxwell gauge theory a complex gauge theory 
or $S^1$ gauge theory. 
Yang-Mills gauge theory, is non-commutative, and has $SU(2)$ gauge symmetry. 
And similarly, since $SU(2)$ is a set of quaternions of norm one, 
that is $S^3$, we may call the Yang-Mills gauge theory 
a quaternionic gauge theory or $S^3$ gauge theory.

Continuing this sequence, we may now arrive at the $S^7$ gauge theory.
$S^7$ is the set of norm one octonions.
With this extension, 
we would now present the construction of non-associative gauge theory, 
using octonions in the place of Lie group.
Before proceeding, we would describe the basics of octonions 
in the following section.

\section{Octonion and Moufang loop}

We first define the notation of octonions. 
$\Bbb{O}=\{x=x^\mu e_\mu \, ; \, x^\mu \in \Bbb{R}\}$ 
is a set of octonions.
$x^\mu$ are real and $e_\mu$ are generators of octonions,
where $e_0=1$ and $e_i \, (i=1,2,\cdots,7)$ are 
pure imaginary units of octonions. 
They satisfy the multiplication rules below~\cite{G-T, DGT},
\begin{eqnarray}
&e_ie_j=-\delta_{ij}e_0+c_{ijk}e_k, \\
&\{e_i,e_j,e_k\}=(e_ie_j)e_k-e_i(e_je_k)
=2f_{ijkl}e_l, \\
&f_{ijkl}=-\frac{1}{3!}\varepsilon_{ijklpqr}c_{pqr}.
\end{eqnarray}
These relation shows that the octonions are 
non-commutative and non-associative.
$c_{ijk}$ represents non-commutativity, 
and $f_{ijkl}$ represents non-associativity.
They also satisfies the relation called the alternative law, 
which assumes associativity only under a special condition.
\begin{eqnarray}
\hspace{-24pt}
a,u \in \Bbb{O},
&&
a(au)=(aa)u, \quad a(ua)=(au)a, \quad u(aa)=(ua)a \\
&&
a(\bar{a}u)=(a\bar{a})u, \quad 
a(u\bar{a})=(au)\bar{a}, \quad 
u(a\bar{a})=(ua)\bar{a} 
\end{eqnarray}
Finally, we also regard the following relations, 
which is called the Moufang identities.
\begin{eqnarray}
a,b,c \in \Bbb{O},\quad 
\left\{
\begin{array}{l}
a(b(ac))=((ab)a)c \\
(ab)(ca)=(a(bc))a=a((bc)a)=a(bc)a \\
b(a(ca))=((ba)c)a
\end{array}
\right.
\end{eqnarray}
Notice that the orders of parentheses are changed.
These identities stand for associativity, 
and we make wide use of them in constructing the theory.
Generally, Moufang loop is defined as the 
alternative division algebra 
which satisfies these Moufang identities~\cite{S-V}.
Octonion is one good example of a Moufang loop.

\section{Octonionic gauge theory} 

\subsection{Gauge potential and gauge field} \label{sec.gaugepot}

Now, we will construct the non-associative gauge theory 
using a Moufang loop as internal symmetry.
We take $S^7$ as Moufang loop.
Ideal introduction would be to define non-associative connection 
and covariant derivative 
by the bundle language of modern mathematics, 
but for convenience and calculational simplicity, 
we present the theory by the formulation 
of Penrose and Rindler~\cite{P-R}.
Following their formulation, 
we take three primitive variables, $\alpha$, $A$, and $F$.
$\alpha$ is called a charged field and is locally
a $S^7$-valued function, 
which we call a {\it gauge}. 
Mathematically, $\alpha$ corresponds to a section of 
``$S^7$ principal bundle'', and physically, 
it represents a reference frame with which we describe the 
{\it charged fields}
\footnote{
One simple physical example of the gauge is 
a reference frame of a rigid body.
To describe a rigid body such as a top in 3-dimensional space, 
one has to first choose a reference axis, 
and then can describe its motion by the $SO(3)$ bases. 
A gauge corresponds to this reference axis.}.
And since $\alpha$ is $S^7$-valued, $\alpha^{-1}=\bar{\alpha}$.
$S^7$ gauge potential $A$ and $S^7$ gauge field $F$
are defined as $A={\alpha}^{-1} \nabla \alpha$ and
$F=\alpha^{-1} \nabla^2 \alpha$.
With Moufang identities, $F$ can be calculated as
\begin{eqnarray}
F&=&dA+(A \alpha^{-1}) \wedge (\alpha A) \label{dAAA}\\
&=&dA + A \wedge A +{\alpha}^{-1}\{\alpha,A, A\}.\label{dAAA2}
\end{eqnarray}
The installation of gauge $\alpha$, is essential.
Notice these $\alpha$ cannot be cancelled because of non-associativity.

\subsection{Gauge transformation and Bianchi identity}

Next, we consider octonionic gauge transformation.
The gauge transformation is defined as usual~\cite{P-R};
$\alpha \to \alpha'=\alpha {g}^{-1}$,
where $g$ is a $S^7$-valued function and $g^{-1}=\bar{g}$. 
Then $A$ and $F$ transforms as
\begin{eqnarray}
A &\to& A'=(g{\alpha}^{-1})((\alpha A) g^{-1})
+(g{\alpha}^{-1})(\alpha dg^{-1}) \\
&&\hspace{16pt}=(g(A \alpha^{-1}))(\alpha g^{-1}) 
-(dg \alpha^{-1})(\alpha g^{-1}),  \\
F &\to& F'= (g{\alpha}^{-1})((\alpha F) g^{-1}) 
=(g(F \alpha^{-1}))(\alpha g^{-1}).
\end{eqnarray}
We can also prove the Bianchi identity for octonionic gauge field,
\begin{eqnarray}
DF&:=&dF + (A{\alpha}^{-1}) \wedge (\alpha F)
-(F\alpha^{-1}) \wedge (\alpha A) \label{Bianchi}\\
&=&dF+[A \wedge F]+\alpha^{-1} \{ \alpha,A,F\}
-\alpha^{-1} \{\alpha,F,A\}\equiv 0. \label{Bianchi2}
\end{eqnarray}
Moufang identities are essential for the calculation.
The triple bracket, $\{\,,\,,\,\}$ represents non-associativity.
This is the difference between the usual gauge theory.

\subsection{Action and equation of motion}

We have prepared the basic mathematical objects 
for the octonionic gauge theory. 
Now we can construct the action and equation of motion.
We take $S^7$ invariant Killing form as,
\begin{eqnarray} 
\langle a,b \rangle:={\rm Re}\{\bar{a}b\}, \label{Killing}
\end{eqnarray}
where $a$ and $b$ are octonions.
And then we have $S^7$, octonionic gauge invariant action as
\begin{eqnarray}
L = \langle \ast F \wedge F \rangle \label{action}
\end{eqnarray}
where $\ast$ is the usual Hodge star, and $F$ is the $S^7$ 
octonionic gauge field defined in section \ref{sec.gaugepot}.
 From this action, we can get the equation of motion for the  
octonionic gauge theory,
\begin{eqnarray}
D \ast F =0 \label{EOM}
\end{eqnarray}
where $D\ast F$ is defined by,
\begin{eqnarray}
D \ast F &=&d\ast F +(A\alpha^{-1}) \wedge (\alpha \ast F)
-(\ast F\alpha^{-1}) \wedge (\alpha A) \\
&=&d \ast F+[A \wedge \ast F]
+\alpha^{-1}\{\alpha,A,\ast F\}
-\alpha^{-1}\{\alpha, \ast F,A\}.
\end{eqnarray}

\subsection{Matter fields}

We also define non-associative matter fields.
It must be reminded that we cannot take a vector space 
as a representation space 
as in the usual gauge theories.
One has to represent non-associative algebra on a Moufang loop.
The $\psi$ is an associated octonionic vector field induced 
from the fundamental representation.
It is a $\Bbb{O}$-valued function, such as
$\psi = \psi^\mu e_\mu$.
According to the octonionic gauge transformation, it transforms as
\begin{eqnarray}
\psi \to \psi'=(g \alpha^{-1})(\alpha \psi)
= g\psi +\alpha^{-1}\{\alpha,g,\psi\}.
\end{eqnarray}
Again, $\alpha$ is installed between $g$ and $\psi$.
Octonionic covariant derivative of $\psi$ 
is defined as,
\begin{eqnarray}
D\psi &:=& d\psi+(A\alpha^{-1}) (\alpha \psi) \\
&=& d\psi +A\psi +\alpha^{-1}\{\alpha,A,\psi\}.
\end{eqnarray}
So we can also prove the Bianchi identity 
(integrablity condition) of
the octonionic matter field, 
\begin{eqnarray}
D^2 \psi = dD\psi+(A\alpha^{-1})(\alpha (D\psi))
=(F \alpha^{-1}) (\alpha \psi).
\end{eqnarray}

\section{BPST like octonionic instanton}

In $\Bbb{R}^8$, there exists BPST like instanton in the
$S^7$ octonionic gauge theory.
First, we represent the point $x$ in the space 
using octonions,
\begin{eqnarray}
x=x^\mu e_\mu, \quad 
\left\{
\begin{array}{l}
e_0=1, \\
e_i \in {\rm Im}\Bbb{O} \quad (i=1,2,\cdots,7).
\end{array}
\right.
\end{eqnarray}
And next, we take a gauge $\alpha$ and gauge potential $A$ such as  
$\displaystyle\alpha=\frac{{x}-{x}_0}{|x-x_0|}$
and $A={\rm Im}\{f(x)d{x}\}$, 
so that the gauge field $F$ becomes self-dual.
$x_0$ is a octonion constant, 
and the function $f(x)$ is given by 
$\displaystyle f(x)=\frac{\bar{x}-\bar{x}_0}{\lambda^2+|x-x_0|^2}$, 
where $\lambda$ is a real constant. 
Self-dual gauge field could be given by 
\begin{eqnarray}
F&=&{\rm Im}\left\{
df \wedge dx +(fdx \cdot \alpha^{-1}) \wedge
(\alpha \cdot fdx)
\right\} \\
&=&
\frac{\lambda^2 e_i}{(\lambda^2+|x-x_0|^2)^2}
\left(dx^0 \wedge dx^i
+\frac12 c_{ijk}dx^j \wedge dx^k\right),
\end{eqnarray}
where self-dual means $\ast F = \frac13 F \wedge \Psi$ in this case~\cite{CDFN}.
$\Psi$ is a special 4-form called Cayley 4-form, and is defined by
\begin{eqnarray}
\Psi=-\frac1{3!}c_{ijk}dx^0 \wedge dx^i \wedge dx^j
\wedge dx^k+\frac1{4!} f_{ijkl} dx^i \wedge dx^j
\wedge dx^k \wedge dx^l.
\end{eqnarray}

\section{Discussion}
We have constructed a non-associative gauge theory 
using Moufang loop as a structure group.
For an example, we presented a $S^7$, octonionic gauge theory.
This octonionic gauge theory is thought to be a natural generalization 
of Maxwell ($S^1$ or $\Bbb{C}$), 
and Yang-Mills ($S^3$ or $\Bbb{H}$) gauge theory.
It has non-associativity by construction, 
and therefore a gauge $\alpha$ remains in the presentation of the theory, 
whereas in the usual associative gauge theory it cancels out naturally. 
For further works, 
one big motivation would be to quantize this theory 
and search the properties of this octonionic gauge theory, 
but even in the classical regime, much application could be considered.
For instance, since this theory is a natural extension 
of associative gauge theory we already know, 
we could expect a rich structure such that could explain 
spontaneous symmetry breaking of internal non-associative symmetry 
to the usual associative symmetry.
Already in classical region, 
we have constructed 7 dimensional non-associative 
Ashtekar gravity using this octonionic gauge theory~\cite{TOL}. 
The application to 8-dimension could be considered as well. 
The spacetime dimension of 7 and 8 are deeply correlated with 
the numbers of the generators of ${\rm Im}{\Bbb O}$.  
While the construction of non-associative Ashtekar gravity, 
we obtained a algebraic structure similar to clifford algebra. 
This indicates that we may construct non-associative Clifford algebra, 
leading to non-associative spinors and non-associative supersymmetry.

\section{Appendix}
Here we give some detailed calculations during the formulation.
We follow the formalism of Penrose and Rindler~\cite{P-R} applied to our 
specific case of Moufang loop.
What we should consider is the difference between 
{\it charged fields} and {\it uncharged fields} which we give a simple explanation below. 
In a word, {\it charged fields} are sections of a bundle,
and {\it uncharged fields} are the usual functions as 
representatives by a local trivialization.
A {\it gauge} $\alpha$ is charged field, $\alpha^{-1}$
is anti-charged field. Gauge potential $A$
and gauge field $F$, are uncharged fields. 
In addition, 
we suppose the existence of covariant derivative of gauge,
$\nabla \alpha$,
and define gauge potential as $A=\alpha^{-1} \nabla \alpha$,
gauge fields as $F=\alpha^{-1} \nabla^2 \alpha$.
It is to be noted that the action of $\nabla$ to uncharged fields 
correspond to taking the exterior derivative, $d$.
First, we show the calculation of octonionic gauge field $F$
by $A$ and $\alpha$, (\ref{dAAA}) and (\ref{dAAA2}).
\begin{eqnarray}
F&=& \alpha^{-1}\nabla (\alpha A)
=\alpha^{-1}(\alpha \nabla A)
+\alpha^{-1} ((\nabla \alpha)A) \\
&=&\alpha^{-1}(\alpha dA) 
+\alpha^{-1} ((\alpha A) \wedge A) \label{F1}
= dA + \alpha^{-1} \left(
\left((\alpha A)(\alpha^{-1}\alpha)\right) \wedge A\right)
\label{totriple}\\ 
&=&dA+\alpha^{-1}
\left((\alpha (A\alpha^{-1})\alpha) \wedge A\right) 
=dA+\alpha^{-1} \left(
\left((\alpha (A\alpha^{-1}))\alpha\right) \wedge A\right) \\
&=&dA+\alpha^{-1} 
\left(\alpha \left((A\alpha^{-1})(\alpha A)\right)\right)
=dA+(A \alpha^{-1}) \wedge (\alpha A). 
\end{eqnarray}
In (\ref{F1}) we inserted $1=\alpha^{-1} \alpha$ between $\alpha A$ 
and $A$, and used Moufang identities.
Intrinsically, $\alpha$ and $\alpha^{-1}$ have only right action 
and left action of octonions respectively.
We can also express $F$ by using the triple brackets. 
 From equation (\ref{totriple}), we get
\begin{eqnarray}
F&=&dA
+\alpha^{-1} ((\alpha A) \wedge A)
=dA+\alpha^{-1} \{\alpha, A, A\}
+\alpha^{-1}\left( \alpha (A \wedge A)\right) \\
&=& dA+A \wedge A +\alpha^{-1} \{\alpha, A, A\}.
\end{eqnarray}
Next we prove octonionic Bianchi identity~(\ref{Bianchi}),
\begin{eqnarray}
dF&=&\nabla(\alpha^{-1} \nabla^2 \alpha)
=\alpha^{-1} \nabla^3 \alpha 
+ (\nabla \alpha^{-1}) \nabla^2 \alpha \\
&=& \alpha^{-1} \nabla^2 (\nabla \alpha)
+(\nabla \alpha^{-1}) \wedge (\alpha F)
=\alpha^{-1} \nabla^2 (\alpha A)
-(A\alpha^{-1}) \wedge (\alpha F) \\
&=& \alpha^{-1} \left( (\nabla^2 \alpha)A+
\alpha d^2 A \right)
-(A\alpha^{-1}) \wedge (\alpha F) \label{dF1} \\
&=&\alpha^{-1}((\alpha F) \wedge A)
-(A\alpha^{-1}) \wedge (\alpha F) \\
&=&\alpha^{-1}\left(((\alpha F)(\alpha^{-1} \alpha)) \wedge A\right)
-(A\alpha^{-1}) \wedge (\alpha F) \\
&=&(F \alpha^{-1}) \wedge (\alpha A)
-(A\alpha^{-1}) \wedge (\alpha F).
\end{eqnarray} 
In (\ref{dF1}) we used nilpotency 
of the exterior derivative, $d^2=0$ 
which is $\nabla^2$ acting on the uncharged field $A$.

Finally, we derive the equation of motion (\ref{EOM}) from 
the action (\ref{action}) in four dimension,
\begin{eqnarray}
\delta L&=&\langle \ast \delta F \wedge F \rangle
+\langle \ast F \wedge \delta F \rangle
=2\langle \ast F \wedge \delta F \rangle \\
&=&  2\left\langle \ast F \wedge 
\left( d\delta A
+\alpha^{-1} ((\alpha A) \wedge \delta A) 
+\alpha^{-1} ((\alpha \delta A) \wedge A) 
\right) \right\rangle \\
&=& 2\left[
\langle \ast F \wedge d\delta A \rangle
+\langle \alpha \ast F \wedge  ((\alpha A) \wedge \delta A) \rangle 
+\langle \alpha \ast F \wedge  ((\alpha \delta A) \wedge A) \rangle
\right] \\
&=& 2\left[
d\langle \ast F \wedge \delta A \rangle
-\langle d\ast F \wedge \delta A \rangle
-\langle ((A \alpha^{-1}) \wedge \alpha \ast F) \wedge 
\delta A \rangle
\right. \nonumber \\ 
&& \left.
+\langle ((\alpha \ast F)\wedge A) \wedge 
(\alpha \delta A) \rangle 
\right] \\
&=&2 d \langle \ast F \wedge \delta A \rangle \nonumber \\
&&-2\left\langle \left(
d \ast F +(A\alpha^{-1}) \wedge (\alpha \ast F)
-(\ast F \alpha^{-1}) \wedge (\alpha A) \right)
\wedge \delta A \right\rangle,
\end{eqnarray}
where we used the fact that $A$ and $F$ are pure imaginary, i.e. 
$\bar{F}=-F$ and $\bar{A}=-A$, and 
the following properties of $S^7$ invariant Killing form. 

For any octonions $a, b, c \in \Bbb{O}$,
\begin{eqnarray}
&\langle a, b \rangle =\langle b, a \rangle, \\
&\langle ab, c \rangle = \langle a,c\bar{b} \rangle
=\langle b, \bar{a}c \rangle.
\end{eqnarray}
This could be derived from the following the octanion's formulas. 
\begin{eqnarray}
&{\rm Re}\{ab\}={\rm Re}\{ba\}=
{\rm Re}\{\bar{a}\bar{b}\}={\rm Re}\{\bar{b}\bar{a}\} \\
&{\rm Re}\{(ab)c\}={\rm Re}\{a(bc)\}.
\end{eqnarray}
The above properties of a Killing form 
denote the $S^7$ invariance of the form (\ref{Killing}).


\end{document}